\documentclass{jkas}

\def\year{2015} 
\def\month{March} 
\def\volume{00} 
\def\issue{0} 
\def\beginpage{1} 
\def\endpage{99} 
\def\received{---} 
\def\accepted{---} 

\date{Received \received ; accepted \accepted}

\title{
Test for Radial Mixing of Stars in M31
}

\author[1]{Andrew Gould}
\author[2]{Hans-Walter Rix}

\affil[1]{Department of Astronomy Ohio State University,
140 W.\ 18th Ave., Columbus, OH 43210, USA 
\email{gould@astronomy.ohio-state.edu }}
\affil[2]{Max-Planck-Institute f\"ur Astronomie, Heidelberg, Germany}

\def\corrauthor{
Andrew Gould
}

\def\runningauthor{
Andrew Gould, Hans-Walter Rix
}

\def\runningtitle{
M31 Radial Mixing
}

\def\keywords{
galaxies: evolution
}

\def\abstracttext{
Effective radial migration and mixing of orbits throughout the stellar disk
has been definitively established in the
Milky Way, but not in any other galaxy.  We show how such radial mixing
can be measured (or strongly constrained) in M31 using a combination
of existing data and readily available facilities.
}

\def\eff{{\rm eff}}

\def\feh{{\rm [Fe/H]}}
\def\apj{{ApJ}}

\def\apjs{{ApJS}}
\def\aap{{A\&A}}

\def\mnras{{MNRAS}}

\begin{document}
\jkashead 


\section{{Introduction: Smoking Gun}
\label{sec:intro}}

\citet{sb02} showed that stars on approximately circular orbits in
disk galaxies are expected to move substantially in galactocentric
radius under the dynamical influence of transient, non-axisymmetric
perturbations in the galactic potential, such as spiral arms or
central bars.

There is a variety of ways to test whether such radial mixing is
present in the Milky Way \citep{sb09}, with one type of observation
acting as a ``smoking gun'' for this effect: disk stars at the solar
circle show a broad range of metallicities at essentially all ages,
while present-day gas-phase metallicities show a very narrow
dispersion at fixed Galactocentric radius and a radial metallicity
gradient (e.g., \citealt{Deharveng2000}).  Together, these two facts
are easily explained by radial mixing of orbits and are essentially
impossible to explain without radial mixing.

If there is radial mixing, then stars that were born at many different
Galactocentric radii (and so in regions with many different gas-phase
metallicities due to the observed gas-phase metallicity gradient)
would all wind up in the solar neighborhood.  Since these stars would
originate in regions of both higher and lower metallicity than the
solar circle, the solar circle would naturally contain a range of
metallicities at each age (except for the very youngest stars that had
not yet mixed).  Indeed the same would be true (and is observed to be
true) at other Galactocentric radii as well.  From the present
perspective, however, this added evidence is just icing on the cake.

On the other hand, the only way to explain this range of metallicities
at fixed age without radial mixing would be to assume that at all
times in the past, the dispersion in gas-phase metallicity at fixed
Galactocentric radius was of order 0.5 dex, and that this dispersion
mysteriously vanished just before our own epoch. Therefore, having
analogous information in even one other comparable galaxy would be
enlightening.

\section{{Does M31 Have Radial Mixing?}
\label{sec:m31_question}}

One would like to compare the strength of radial mixing in the Milky
Way with that in other galaxies.  From theory, one expects that in a
purely axisymmetric galaxy there would be no radial mixing.  By
extension, more weakly axisymmetric galaxies should have weaker
mixing.  M31 is the obvious choice for the first comparison because it
is both nearby and appears to have substantially weaker
non-axisymmetries than the Milky Way.  To our knowledge there have
been no direct tests of radial mixing in M31.  However, there is one
hint that it may be weaker: the bright ring at 10 kpc has an
over-density of stars ranging from 100 Myr to 500 Myr, but the older
stars are not noticeably more diffuse in radius
\citep{m31ring}. However, numerical simulations indicate that the
effects of radial migration only become manifest after a number of
local orbital periods, i.e., beyond 500~Myrs \citep{Minchev2011}.

How strongly radial migration is manifested in a metallicity spread at
a given stellar age depends on the radial gradient of the gas
metallicity. \citet{Yin09} have compared gas-phase (or young star)
abundance gradients in the Milky Way and M31. There is still
considerable debate about the M31 metallicity gradient
\citep{Trundle2002,Sanders2012}: the gradient is presumably somewhat
shallower than that of the Milky Way, but a number of tracers indicate
a small abundance spread in very young stars at a given radius
\citep{Trundle2002}.

Recent observations of the age -- velocity dispersion relation in
M31's stellar disk \citep{Dorman2015} show that this relation differs
markedly between the Milky Way and M31: at a given age, M31's disk
stars have a higher velocity dispersion, and their velocity dispersion
depends more strongly on age than in the Galaxy. This points toward a
different disk formation history in M31, and it would be important to
understand whether these differences also translate into a difference in
radial migration.  Recent numerical simulations suggest that these
higher velocity dispersions should make radial migration far less
efficient \citep{Vera-Ciro2014}.

\section{{Proposed Test for M31 Radial Mixing}
\label{sec:m31_answer}}

The ``smoking gun'' test for effective radial migration and mixing
that was outlined in Section~\ref{sec:intro} can be directly applied
to M31.  For stars at or above the main-sequence turnoff, ages and
metallicities can be measured by combining determinations of the
effective temperature $T_\eff$, luminosity $L$ and metallicity $\feh$.
Even with 3-band photometry, one can solve for $T_\eff$, single-band
flux $F$, and extinction $A_V$ under the assumption of known $\feh$
and extinction-law $R_V$.  With four bands, one can solve for the
extinction law as well.  Presently high-resolution 6-band photometry
is available from the PHAT survey \citep{phat}.  By adding in
moderate-resolution (or even low-resolution) spectroscopy, one can
solve simultaneously for $(T_\eff, L, \feh,A_V, R_V, F)$.  Given that
all stars in M31 are at a common (and reasonably well known) distance,
one can derive $L$ from $(F,T_\eff,\feh)$. The viability of such
determinations has recently been demonstrated \citep{Dorman2015}.

Moderate-resolution optical spectrographs on large telescopes are
already available (e.g., LRIS on Keck or MODS on LBT, \citealt{mods})
and even more powerful ones will be available within a few years.
With \feh\ obtained via low- or moderate-resolution spectroscopy one
could therefore measure ages and metallicities on thousands of M31
giants down to several magnitudes below the tip of the red giant
branch.  If spectroscopy focuses on the rare stars near the tip of the
giant branch, not only does it become easier to get sufficient S/N,
but source crowding becomes less detrimental.

\section{{From Existence-Proof to Probe}
\label{sec:probe}}

In practice, two tests to quantify the efficacy of radial migration in
M31 seem feasible.  The first is mapping the metallicity distribution
at a given stellar age, $p(\feh \mid \tau_{age})$: in the presence of
a spatial $\feh$ gradient, the dispersion of this distribution should
go up with age. The mean of the age-metallicity relation will depend
on the radial-age and radial-metallicity gradients.

A second type of test can come from the shape of the metallicity
distribution function: in the case of the Milky Way, the metallicity
distribution is approximately Gaussian in the solar neighborhood but
is skewed both further out and further in (in opposite senses, R.\
Sch\"onrich 2014, private communication).  Detailed modeling of these
distributions (including as a function of age) provides a window into
the history of our Galaxy.  If radial mixing is detected in M31, the
next step would be to measure it quantitatively.  Modeling of these
distributions would then enable a comparison of Milky Way and M31
histories, as well as probing for links between the degree of radial
mixing with the strength and form of non-axisymmetries in each galaxy.

Analyzing M31 in this way, would provide the most immediate check on
the seemingly high efficacy of radial migration inferred for the Milky
Way.  M33 is another spiral galaxy, with a far weaker bulge but with a
grand design spiral pattern \citep{Regan94}, for which multi-band HST
imaging exists, and spectroscopy from the tip of the giant branch
would provide the metallicities and ages for an analogous test.

With the next generation of telescopes, either ground-based 30m or
JWST, galaxies like M81, NGC300, and others may also be available to
test whether radial migration is indeed the mechanism that determines
the present-day galactocentric radii of individual stars.


\acknowledgments

Work by AG was supported by NSF grant AST 1103471.
HWR's research leading to these results has received funding from the
European Research Council under the FP 7  ERC Grant Agreement n.~[321035].
This research was inspired by the interactive environment at the
Galactic Archaeology Workshop at the Kavli Institute for Theoretical Physics
in Santa Barbara.

\end{document}